\documentclass[twocolumn,showpacs,preprintnumbers,amsmath,amssymb,showkeys]{revtex4}
\usepackage{graphicx}% Include figure files
\usepackage{dcolumn}% Align table columns on decimal point
\usepackage{bm}% bold math
\usepackage{longtable}

\begin{document}

\preprint{APS/123-QED}

\title{Measurement of the neutron capture cross section of the $s$-only isotope
  $^{204}$Pb from 1~eV to 440~keV}

\author{
C.~Domingo-Pardo$^{1,2}$\footnote{Corresponding author. E-mail: cesar.domingo.pardo@cern.ch.},
U.~Abbondanno$^{3}$, 
G.~Aerts$^{4}$, 
H.~\'Alvarez-Pol$^{5}$, 
F.~Alvarez-Velarde$^{6}$, 
S.~Andriamonje$^{4}$, 
J.~Andrzejewski$^{7}$, 
P.~Assimakopoulos $^{8}$,    
L.~Audouin $^{1}$, 
G.~Badurek$^{9}$
% ...
\begin{center}
The n\_TOF Collaboration
\end{center}}

\affiliation{
{$^{1}$}Forschungszentrum Karlsruhe GmbH (FZK), Institut f\"{u}r Kernphysik, Germany;\,  
{$^{2}$}Instituto de F{\'{\i}}sica Corpuscular, CSIC-Universidad de Valencia, Spain;\, 
{$^{3}$}Istituto Nazionale di Fisica Nucleare, Trieste, Italy;\,  
{$^{4}$}CEA/Saclay - DSM, Gif-sur-Yvette, France;\,   
{$^{5}$}Universidade de Santiago de Compostela, Spain;\,   
{$^{6}$}Centro de Investigaciones Energeticas Medioambientales y Technologicas, Madrid, Spain;\,   
{$^{7}$}University of Lodz, Lodz, Poland;\,  
{$^{8}$}University of Ioannina, Greece;\,   
{$^{9}$}Atominstitut der \"{O}sterreichischen Universit\"{a}ten,Technische Universit\"{a}t Wien, Austria;\,\,\,  
% ...
}

%\email{Cesar.Domingo.Pardo@cern.ch}

\date{\today}%

\begin{abstract}
The neutron capture cross section of $^{204}$Pb has been measured at the
CERN n\_TOF installation with high resolution in the energy range from
1~eV to 440~keV. An R-matrix analysis of the resolved resonance region,
between 1~eV and 100~keV, was carried out using the SAMMY code. In
the interval between 100~keV and 440~keV we report the average
capture cross section. The background in the entire neutron energy range
could be reliably determined from the measurement of a $^{208}$Pb
sample. Other systematic effects in this measurement could be investigated
and precisely corrected by means of detailed Monte Carlo simulations. We
obtain a Maxwellian average capture cross section for $^{204}$Pb at
$kT=30$~keV of 79(3)~mb, in agreement with previous experiments. However our
cross section at $kT=5$~keV is about 35\% larger than the values reported so
far. The implications of the new cross section for the $s$-process abundance
contributions in the Pb/Bi region are discussed.
\end{abstract}

\maketitle

\section{\label{sec:intro}Introduction}

The heaviest stable isotopes with masses $A=204$ - $209$ are 
synthesized by neutron capture reactions, the $s$ and the 
$r$ processes. According to the stellar model of Arlandini et
al.~\cite{arl99}, the $s$-process fraction of $^{204,206}$Pb is mostly
produced in thermally pulsing asymptotic giant branch (\textsc{agb}) stars,
the so called main component of the $s$ process. 
On the other hand, the galactic chemical evolution study of Travaglio et
al.~\cite{tra99,tra01} showed that the heavier lead isotopes $^{207,208}$Pb
and bismuth are basically synthesized by early generation, low-metallicity,
low-mass \textsc{agb} stars.
Bismuth is the last element synthesized by the slow process, thus further
neutron captures on this isotope are recycled back to $^{206,207,208}$Pb via
$\alpha$-decays.

The situation at the end of the $s$ process is complicated due 
to branchings in the $\alpha$-recycling at $^{210}$Po ($t_{1/2} 
= 138$~d) and at $^{210m}$Bi ($t_{1/2} = 3$~Myr). In this 
termination region, $^{204}$Pb is the only of pure $s$-process
origin, because it is shielded from the $r$ process by its isobar
$^{204}$Hg. Therefore, $^{204}$Pb is important for disentangling 
the complex Pb/Bi abundance pattern. The solar abundance and the 
cross section of $^{204}$Pb need to be accurately known for a 
consistent determination of the $s$-process components of the 
Pb/Bi abundances, which provides a basis for constraining the 
complementary contributions from explosive $r$-process 
nucleosynthesis. 

With an improved $s$-process part, the respective 
$r$ components, which consist of the direct $r$-process yields as 
well as of the decay products from the $\alpha$-unstable trans-bismuth 
region, could be more accurately determined \cite{rat04}. The 
radiogenic fractions are important in order to consolidate the 
validity of the U/Th cosmochronometer~\cite{cow91,cow99,sch02,kra04}.
The cross section of $^{204}$Pb also enters into the calculation of 
the $s$-process branching at $^{204}$Tl. Since this branching shows 
a strong temperature dependence, the abundance of $^{204}$Pb represents 
an important test for \textsc{agb} models, which exhibit strongly
different neutron densities and temperatures in and between thermal 
pulses \cite{GAB98}. 

Thanks to improvements both in experimental techniques and detectors, 
difficulties in previous measurements of the ($n, \gamma$) cross
section of $^{204}$Pb \cite{hor84} could be significantly reduced. This 
concerns the investigated neutron energy range, which had been covered
only for energies above 2.5~keV with the consequence that some
important resonances were missed. It also concerns the correction for
background from neutrons scattered in the sample, which had a 
strong effect on the capture width of broad resonances. The setup in
previous experiments suffered from large scattering corrections with 
uncertainties of $\sim$50\%. Apart from this problem, the remaining 
systematic uncertainties had been estimated to be $\pm$5\%~\cite{hor84}.

The ($n, \gamma$) cross section measurement at the CERN n\_TOF
facility ~\cite{abb03} has covered the full energy range between 1~eV 
and 1~MeV in a single experiment, and  the corrections due to scattered 
neutrons became negligible for all resonances by using C$_6$D$_6$ 
detectors with reduced neutron sensitivity~\cite{pla03}. Furtheron, 
systematic uncertainties were improved to the level of 3\%~\cite{abb04}
by detailed Monte Carlo simulations of the experimental setup.

\section{\label{sec:CS measurement} Cross Section measurement}

The present measurement was carried out with a $^{204}$Pb sample of
99.7\% isotopic enrichment. At n\_TOF, neutrons are produced by 
spallation reactions using a pulsed proton beam (6~ns (rms), 20~GeV/$c$)
impinging on a lead block. A water layer around the lead target 
serves as moderator of the initially fast neutron spectrum, as well 
as coolant of the spallation target. Particularly relevant for this 
measurement was the low n\_TOF duty cycle with a pulse repetition rate
of 0.4~Hz, which allows us to cover a wide energy range from 1~MeV 
down to 1~eV. A further advantage of the present measurement is the 
small sample thickness of $n=0.00376$~at/barn, more than 7 times
thinner compared to the sample used in a previous measurement
\cite{hor84,all73}. In this way, systematic effects due to multiple 
scattering and neutron self absorption in the sample become rather low. 

The sample was mounted on the ladder of an evacuated sample changer 
made from carbon fiber. In addition a thin gold sample was also 
regularly measured for absolute yield normalization via the saturated 
resonance technique~\cite{mac79}, and an enriched $^{208}$Pb sample,
which has a negligibly small ($n, \gamma$) cross section with only 
few resonances in the investigated energy range, served for the 
the determination of the in-beam $\gamma$-ray background produced 
by neutron captures in the water moderator of the lead spallation 
target. Due to the relatively large cross section of $^{204}$Pb, 
this background was only a minor difficulty for the present 
measurement.  

\begin{table}[htbp]
\caption{\label{tab:samples} Sample characteristics$^a$}
\begin{ruledtabular}
\begin{tabular}{lccc}
Sample     & Mass & Thickness  & Isotopic composition \\
           & (g) & (at/barn) &  (\%)\\

\hline
$^{204}$Pb &   4.039   & 0.00376  & 99.7 \\
$^{208}$Pb &   12.53   & 0.01155  &  99.86    \\
$^{197}$Au &   0.768   & 0.00074  &  100   \\
\end{tabular}
\end{ruledtabular}
$^a$ All samples were 20 mm in diameter.
\end{table}

Neutron capture events were registered via the prompt capture 
$\gamma$-ray cascade by a set of two C$_6$D$_6$ detectors, which 
were optimized with respect to neutron sensitivity~\cite{pla03}. 
The detectors were placed at 125$^{\circ}$ with respect to the 
direction of the neutron beam in order to minimize angular 
distribution effects as well as the background due to in-beam
$\gamma$-rays. A schematic view of the experimental setup can be
seen in Fig.~2 of Ref.~\cite{dom06}. The neutron flux $\Phi_n(E_n)$ was previously
determined by measuring the well known $^{235,238}$U fission 
yields~\cite{ptb}. During the experiment it was determined
by means of the saturated gold resonance at 4.9 eV measured with 
the gold sample, and it was also monitored by means of a 
monitor detector consisting of a thin $^6$Li foil surrounded
by a set of four silicon large detectors for recording the 
products of the $^{6}$Li($n, \alpha$)$^3$H reactions~\cite{mar04}.

\section{\label{sec:analysis}Data analysis}

Since the $\gamma$-ray efficiency of the C$_6$D$_6$ detectors 
is rather small, their response function needs to be appropriately 
weighted in order to achieve a cascade detection probability
independent of the particular $\gamma$-ray registered. This is 
achieved by applying the pulse height weighting technique (PHWT)
\cite{mac67}. In the present analysis the weighting functions 
(WF) for the measured lead and gold samples were obtained via 
the Monte Carlo technique, following the procedure described in
Refs.~\cite{abb04,dom06}.
 
The experimental capture yield $Y^{exp}$ can then be determined 
from the measured and weighted count rate ($N^w$),
\begin{equation}\label{eq:yexp}
Y^{exp}(E_n) = f^t\, f^{sat} \, \frac{N^w(E_n)}{\Phi_n(E_n) \, E_c(E_n)},
\end{equation}
where $E_c$ is the neutron capture energy, $f^{sat}$ an absolute 
yield normalization factor determined from the analysis of the 
4.9~eV saturated resonance in the gold runs, and $f^t$ is a yield 
correction factor, which accounts for the effect of the threshold 
in the pulse height spectra of the C$_6$D$_6$ detectors. The 
latter corrections, which were obtained by Monte Carlo simulations 
as described in Refs.~\cite{abb04,dom06}, were found to be 3.1(3)\% 
for resonances with spin $J=1/2$ and to 3.6(3)\% for $J=3/2$ 
resonances. The treatment of the experimental background will be 
described in the two following sections.

The systematic uncertainties of the present measurement are 
summarized in Table~\ref{tab:uncertainty}.

\begin{table}[htbp]
\caption{\label{tab:uncertainty} Systematic uncertainties in the
  measured cross section of $^{204}$Pb.}
\begin{ruledtabular}
\begin{tabular}{lc}
Effect & Uncertainty (\%)\\
\hline
PHWT and yield normalization factors, $f^{sat}$ & $<2$      \\
Background subtraction                         & 1(10)$^a$ \\
Flux shape                                     & 2         \\
Yield correction factors, $f^t$                & 0.3       \\
                                               &           \\
Total systematic uncertainty                   & 3 (10)$^a$\\
\end{tabular}
\end{ruledtabular}
$^a$ Values in brackets refer to the unresolved resonance 
     region between 100 and 440~keV.
\end{table}

\section{\label{sec:results RRR} Results in the resolved resonance region}

In the resolved resonance region (\textsc{rrr}), the experimental 
yield (\ref{eq:yexp}) is described by means of the R-matrix 
formalism in terms of individual resonance parameters using an 
equation of the type, 

\begin{equation}\label{eq:yield}
Y^{exp} = B(E_n) + Y(E_{\circ}, \Gamma_n,\Gamma_{\gamma}).
\end{equation}

Where available, the neutron widths $\Gamma_n$ from literature
\cite{mug06} have been used as input for the present analysis. The 
capture width $\Gamma_{\gamma}$ of each observed resonance was 
fitted with the R-matrix code \textsc{SAMMY}~\cite{lar06}, which
includes also corrections for several experimental effects, e.g.
for Doppler broadening, multiple neutron scattering and self 
shielding in the sample. The background term $B(E_n)$ could be 
precisely determined from the concomitant ($n, \gamma$) measurement 
with a $^{208}$Pb sample. Given the much lower capture cross 
section of $^{208}$Pb, the C$_6$D$_6$ response function to 
in-beam $\gamma$-rays scattered by the $^{204}$Pb sample could be 
directly determined from the measured $^{208}$Pb spectrum. The 
contribution from scattered $\gamma$-rays dominated the overall 
background in the present measurement by far. 

In the interval from 1~eV to 30~keV,
$B(E_n)$ could be adjusted to a function of the type,

\begin{equation}
B(E_n) = A_1 + \frac{A_2}{\sqrt{E_n}} + A_3 \sqrt{E_n}.
\end{equation}

\begin{figure}[h]
\includegraphics[width=0.45\textwidth]{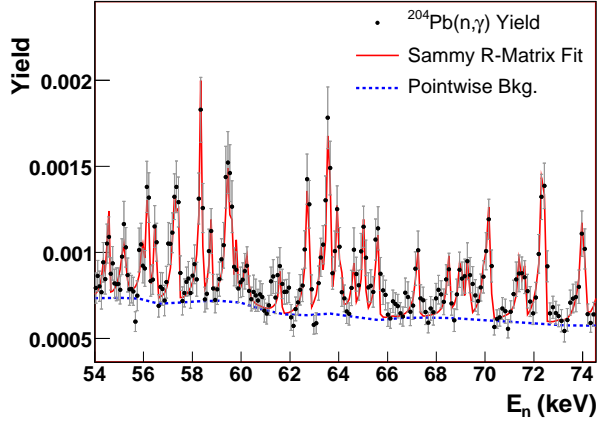}
\caption{\label{fig:bkg} (Color online) $^{204}$Pb capture 
yield and pointwise background in the neutron energy region between
54 and 74~keV.}
\end{figure}

Between 30 and 100~keV the background showed systematic
fluctuations, which could not be described by means of a 
single analytical function. Hence, the background was 
defined in that energy range by a pointwise numerical 
function, as illustrated in Fig.~\ref{fig:bkg}.

The capture widths, $\Gamma_{\gamma}$, obtained in this analysis 
are listed in Table~\ref{tab:RP}. Also, the capture kernels 
\begin{equation}
K_r = \frac{2J+1}{2} \frac{\Gamma_{\gamma}\Gamma_n}{\Gamma_{\gamma}+\Gamma_n}.
\end{equation}

are given for each case together with the respective uncertainties.

\begin{longtable}{cccccccc}
\caption{Resonance parameters derived from the R-matrix analysis of the
  $^{204}$Pb($n, \gamma$) data.}\label{tab:RP}\\
\hline
\hline
$E_{\circ}$ & $l$ & $J$ & $\Gamma_{\gamma}$ & $\Delta{\Gamma_{\gamma}}$ & $\Gamma_{n}$ & $K_r$ & $\Delta{K_r}$\\
(eV)   &   &     & (meV)  & (\%)   & (meV)   & (meV) & (\%)  \\
\hline
\hline
480.3  &  1 & 1/2 &  1.33 & 4 &  3.0  &  0.92$^a$ & 2.7 \\
1333.8  &  1 & 1/2 &  105 & 4 &  46.3$^b$  &  32.1$^a$ & 1.3 \\
1687.1  &  0 & 1/2 &  1029 & 0.7 &  3340  &  787$^a$ & 0.5 \\
2481.0  &  0 & 1/2 &  514 & 1.1 &  5470  &  470$^a$ & 1.0 \\
2600.0  &   &  &   &  &   &  8.35 & 6 \\
2707.1  &  1 & 3/2 &  31.2 & 9 &  11.5  &  16.8 & 2 \\
3187.9  &  0 & 1/2 &  316 & 10 &  1.7  &  1.69 & 0.1 \\
3804.9  &  1 & 1/2 &  280 & 8 &  66.4  &  53.7 & 1.6 \\
4284.1  &  1 & 3/2 &  111 & 9 &  24.0  &  39.4 & 1.7 \\
4647.5  &   &  &   &  &   &  2.57 & 9 \\
4719.4  &  1 & 3/2 &  41.2 & 5 &  95.0  &  57.5 & 3 \\
5473.2  &  1 & 1/2 &   &  &    &  79.0 & 1.6 \\
5561.4  &   & (1/2) &  1.03 & 10 &  1.9  &  0.67 & 6.4 \\
6700.5  &  0 & 1/2 &  312 & 3 &  4540  &  292 & 3 \\
7491.0  &   &  &   &  &    &  19.0 & 0.5 \\
8357.4  &  0 & 1/2 &  1286 & 1.9 &  45000  &  1250 & 1.9 \\
8422.9  &   &  &   &  &   &  11.3 & 7 \\
8949.6  &   &  &   &  &   &  22.9 & 3 \\
9101.0  &   & (1/2) &  193 & 8 &  150  &  84.4 & 4 \\
9649.3  &  0 & 1/2 &  1076 & 2 &  7860  &  946 & 2 \\
10254  &   &  &   &  &  &  37.0 & 8 \\
11366  &  1 & 3/2 &  39.0 & 10 &  226  &  66.5 & 9 \\
11722  &   &  &   &  &  &  22.8 & 9 \\
12147  &   &  &   &  &  &  54.4 & 8 \\
12519  &   &  &   &  &  &  24.3 & 9 \\
12909  &  0 & 1/2 &  569 & 4 &  54600  &  563 & 4 \\
13007  &   &  &   &  &   &  6.07 & 10 \\
13382  &  1 & 3/2 &  55.1 & 10 &  232  &  89.0 & 8 \\
14377  &   &  &  & &  &  47.8 & 10 \\
14822  &  0 & 1/2 &  548 & 4 &  4301  &  486 & 4 \\
15947  &  1 & 1/2 &  201 & 10 &  130  &  79.0 & 4 \\
16077  &   &  &  &  &   &  16.2 & 10 \\
16121  &   &  &  &  &   &  66.0 & 8 \\
16493  &   &  &  &  &   &  19.9 & 10 \\
17433  &   &  &  &  &   &  39.3 & 9 \\
17455  &  1 & 1/2 &  528 & 9 &  260  &  174 & 3 \\
17647  &  1 & 3/2 &  62 & 0.0 &  440  &  109 & 0.0 \\
18092  &   &  &   &  &  &  31.7 & 9 \\
18299  &   &  &   &  &  &  19.1 & 10 \\
18511  &  1 & 1/2 &  362 & 9 &  259  &  151 & 4 \\
18597  &  &  &  &  &  &  10.1 & 10 \\
18677  &  &  &  &  &  &  9.52 & 10 \\
18806  &  0 & 1/2 &  81.3 & 9 &  230  &  60.0 & 7 \\
19748  &  0 & 1/2 &  738 & 6 &  2530  &  571 & 4 \\
20396  &  1 & 1/2 &   &  &   &  92.6 & 8 \\
20776  &  1 & 1/2 &  202 & 9 &  300  &  121 & 5 \\
20979  &   & &  &  &  &  41.0 & 9 \\
21178  &   & &  &  &  &  43.1 & 9 \\
21659  &  1 & 1/2 &  258 & 8 &  630  &  183 & 6 \\
22061  &  &  &  &  &  &  77.1 & 9 \\
22209  &  0 & 1/2 &  463 & 6 &  56833  &  459 & 6 \\
23031  &  & &  & &  &  21.7 & 10 \\
23290  &  1 & 3/2 &  99.0 & 10 &  1245  &  183 & 9 \\
23379  &  & & & & &  55.1 & 9 \\
23968  &  & & & & &  111 & 8 \\
24158  &  0 & 1/2 &  126 & 10 &  77300  &  126 & 10 \\
24184  &  & & & &  &  124 & 10 \\
24510  &  & (1/2) &  73.0 & 10 &  450  &  62.8 & 8 \\
25446  &  & &  & &  &  118 & 8 \\
25711  &  & &  & &  &  117 & 8 \\
25805  &  & &  & &  & 76.6 & 9 \\
25914  &  1 & 1/2 &  75.7 & 10 &  710  &  68.4 & 9 \\
26241  &  &  & &  &  &  171 & 9 \\
26665  &  &  & &  &  &  83.2 & 9 \\
27207  &  &  & &  &  &  90.2 & 9 \\
27410  &  &  & &  &  &  200 & 7 \\
27590  &  0 & 1/2 &  747 & 6 &  30300  &  729 & 6 \\
27884  &  0 & 1/2 &  429 & 7 &  6162  &  401 & 7 \\
28144  &  1 & 1/2 &  129 & 9 &  950  &  114 & 8 \\
28950  &   & (1/2) &  179 & 10 &  330  &  116 & 6 \\
29043  &  1 & 1/2 &  100 & 9 &  1040  &  91.6 & 8 \\
29222  &   &  &  &  &  &  87.1 & 9 \\
29565  &  &  &   &  &  &  84.5 & 9 \\
29671  &  1 & 1/2 &  185 & 9 &  1250  &  161 & 8 \\
30302  &  &  &  &  &  &  220 & 7 \\
31200  &  &  &  &  &  &  90.0 & 9 \\
31487  &  & (1/2) &  276 & 10 &  300  &  144 & 5 \\
32647  & & & & & &  348 & 6 \\
32853  &  0 & 1/2 &  781 & 7 &  43934  &  767 & 7 \\
33504  &  1 & 1/2 &  144 & 10 &  1360  &  130 & 9 \\
33708  &  1 & 1/2 &  47.7 & 10 &  1000  &  45.5 & 10 \\
33946  &  0 & 1/2 &  448 & 9 &  1380  &  338 & 7 \\
34234  &  1 & 3/2 &  81.0 & 9 &  8268  &  160 & 9 \\
35696  &  & & & & &  200 & 8 \\
35981  &  & & & & &  267 & 7 \\
36797  &  1 & 1/2 &  30.0 & 10 &  4360  &  29.8 & 10 \\
37720  &  1 & 3/2 &  103 & 10 &  325  &  156 & 7\\
38455  &  &  &  & & & 123 & 8 \\
38732  &  1 & 3/2 &  52 & 9 &  855  & 98.5 & 9\\
38977  &  1 & 1/2 &  230 & 9 &  1840  &  204 & 8 \\
39557  &  0 & 1/2 &  1361 & 6 &  158000  &  1349 & 6  \\
39890  &  1 & 1/2 &  131 & 10 &  2780  &  125 & 9 \\
40520  &  1 & 1/2 &  250 & 9 &  1230  &   207 & 8  \\
40888  &  1 & 3/2 &  270 & 9 &  545  &  361 & 6  \\
41670  &  0 & 1/2 &  221 & 9 &  7050  &   214 & 9\\
42380  &  1 & 3/2 &  182 & 9 &  1835  &  331 & 8  \\
42496  &  1 & 1/2 &  58 & 10 &  7580  &   57.8 & 10  \\
42962  &  0 & 1/2 &  570 & 8 &  46300  &  563 & 8 \\
43080  &  1 & 3/2 &  70 & 10 &  2590  &   136 & 10   \\
43725  &  1 & 1/2 &  189 & 10 &  2670  &  177 & 9 \\
43938  &  1 & 3/2 &  2520 & 10 &  355  &   622 & 1.2 \\
44471  &  1 & 1/2 &  170 & 10 &  920  &  144 & 8 \\
44950  &  0 & 1/2 &  467 & 9 &  21409  &  457 & 8 \\
45370  &  0 & 1/2 &  353 & 9 &  2780  &  313 & 8 \\
45527  &   &  &  & &  &  406 & 8 \\
45886  &  1 & 1/2 &  107 & 10 &  6220  &  105 & 10 \\
46263  &  1 & 3/2 &  283 & 8 &  2885  &  515 & 8 \\
46700  &  & & & & &  152 & 9 \\
47453  &  & & & & &  237 & 9 \\
47880  &  1 & 3/2 &  212 & 9 &  2520  &  392 & 8 \\
49306  &  0 & 1/2 &  453 & 9 &  59400 &   622 & 1.2  \\
50229  &  1 & 1/2 &  221 & 10 &  1900  &   198 & 9  \\
50490  &  & & & & &  215 & 8 \\
50827  &  & & & & &  321 & 7 \\
51250  &  & & & & &  46.5 & 10 \\
51581  &  1 & 1/2 &  73.6 & 10 &  3260  &  71.9 & 10 \\
52809  &  1 & 3/2 &  158 & 9 &  510  &  241 & 7 \\
54260  &  1 & 1/2 &  75.0 & 10 &  1880  &  72.1 & 10 \\
54476  &  1 & 1/2 &  352 & 9 &  3251  &  318 & 8 \\
55118  &  0 & 1/2 &  420 & 9 &  153000  &  419 & 9 \\
55857  &  1 & 1/2 &  163 & 10 &  2260  &  152 & 9 \\
56084  &  1 & 3/2 &  428 & 10 &  500  &  461 & 5 \\
56397  &   & (1/2) &  431 & 9 &  1460  &  333 & 7 \\
57179  &  0 & 1/2 &  392 & 9 &  36100  &  388 & 9 \\
57310  &   &  &  & &  &  450 & 5 \\
58266  &  1 & 3/2 &  686 & 9 &  1300  &  898 & 6  \\
58674  &   & & & & &  302 & 8 \\
59334  &  0 & 1/2 &  548 & 9 &  4900  &  493 & 8 \\
59491  &  1 & (1/2) &  1029 & 10 &  820  &  456 & 4 \\
59717  &  1 & 3/2 &  123 & 10 &  4467  &  239 & 9 \\
60135  &  0 & 1/2 &  312 & 10 &  93980  &  311 & 10 \\
61500  &   & & & & &  196 & 8 \\
62635  & 1 & 3/2 & 453 & 9 & 1000 & 624 & 6 \\
62648  &   & (1/2) & 100 & 10 &  350  & 78.5 & 8\\
63156  &   & (1/2) &  212 & 10 &  2180  &  194 & 9 \\
63492  &   & (1/2) &  2690 & 9 &  1770  &  1068 & 4  \\
63854  &   & (3/2) &  275 & 9 &  6864  &  529 & 9 \\
64002  &  1 & 1/2 &  213 & 10 &  2100  &  193 & 9 \\
64500  &  &  & & & &  231 & 8 \\
64925  &  &  & & & &  502 & 5 \\
65480  &  0 & 1/2 &  570 & 9 &  10000  &  540 & 8 \\
67122  &  0 & 1/2 &  444 & 9 &  13400  &  429 & 9 \\
68395  &  & & & & &  333 & 6 \\
68882  &  & & & & &  369 & 6 \\
69191  &  & & & & &  383 & 6 \\
69870  &  1 & 3/2 &  134 & 10 &  10000  &  265 & 10 \\
70055  &  1 & 1/2 &  2494& 10 &  1000  &   714 & 3  \\
71294  &   &  &  &  &  &  328 & 7 \\
71477  &   &  &  &  &  &  367 & 6 \\
72249  &  1 & 3/2 &  1136 & 9 &  1755  &   1379 & 5 \\
73885  &  0 & 1/2 &  963 & 8 &  41800  &  941 & 8 \\
74683  &  0 & 1/2 &  721 & 9 &  149990  &  718 & 9 \\
75456  &   &  &  & &  &  995 & 0.1 \\
78323  &  0 & 1/2 &  399 & 10 &  68001  &  397 & 9 \\
79547  &  1 & 3/2 &  172 & 10 &  8400  &  338 & 9 \\
80540  &  0 & 1/2 &  1268 & 8 &  64015  &  1244 & 8 \\
82256  &  0 & 1/2 &  989 & 9 &  55707  &  972 & 8 \\
83940  &  1 & 3/2 &  137 & 10 &  19700  &  271 & 10 \\
84334  &  0 & 1/2 &  1033 & 9 &  8480  &  921 & 8 \\
84980  &   & (1/2) &  988 & 9 &  4930  &  823 & 8 \\
86013  &   & (1/2) &  1074 & 9 &  3000  &  791 & 7 \\
86765  &  0 & 1/2 &  1554 & 9 &  258129  &  1545 & 8 \\
88071  &  0 & 1/2 &  1058 & 9 &  43200  &  1033 & 9 \\
89052  &  1 & 3/2 &  447 & 9 &  21202  &  875 & 9 \\
90794  &  0 & 1/2 &  1348 & 9 &  26700  &  1283 & 8 \\
91530  &  1 & 3/2 &  941 & 9 &  2000  &  1279 & 6 \\
92323  &  0 & 1/2 &  531 & 10 &  143985  &  529 & 10 \\
93561  &  0 & 1/2 &  505 & 10 &  136012  &  503 & 9 \\
95080  &  2 & 3/2 &  982 & 9 &  6601  &  1710 & 7 \\
96298  &  1 & 3/2 &  1021 & 8 &  7850  &  1808 & 7 \\
98123  &  1 & 3/2 &  274 & 10 &  23351  &  543 & 9 \\
\hline
\multicolumn{8}{c}{$^{a}$ First determination in a capture experiment.}\\
\multicolumn{8}{c}{$^{b}$ Neutron width fitted as $\Gamma_n=$46.3$\pm$2.5~meV.}\\
\end{longtable}

\section{\label{sec:results URR} Results in the unresolved resonance region}

\begin{figure}[h]
\includegraphics[width=0.4\textwidth]{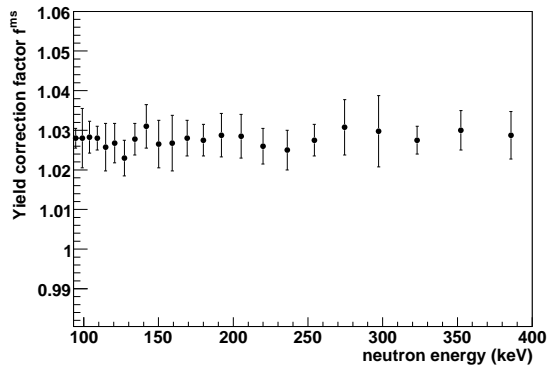}
\caption{\label{fig:fms} Correction factor $f^{ms}(E_n)$ due to self-absorption
  and multiple scattering calculated with the code \textsc{sesh}~\cite{fro68}.}
\end{figure}

The average capture yield $\left<Y(E_n)\right>$ is related to 
the average capture cross section $\left< \sigma_{\gamma}(E_n)\right>$ 
by
\begin{equation}
\left<Y(E_n)\right> = n\, f^{ms}(E_n)\, \left< \sigma_{\gamma}(E_n)\right>,
\end{equation}
where $n$ is the sample thickness in atoms per barn and $f^{ms}(E_n)$ 
is the neutron self-shielding and multiple scattering correction. This 
correction was determined via the Monte Carlo technique using
the code \textsc{sesh}~\cite{fro68}. In the considered region between
100 and 400 keV the correction factors $f^{ms}(E_n)$ are practically
constant as shown in Fig.~\ref{fig:fms}. 

The averaged cross sections $\left<\sigma_{\gamma}(E_n)\right>$ are
given in Table~\ref{tab:URR} together with the respective statistical 
uncertainties. An overall systematic uncertainty of $\pm$10\% has to be
added in order to account for the systematic uncertainties of $f^{ms}(E_n)$
and of the background subtraction in this energy range.

\begin{table}[htbp]
\caption{\label{tab:URR} Average neutron capture cross section for $^{204}$Pb.}
\begin{ruledtabular}
\begin{tabular}{cccc}
$E_{\rm low}$ & $E_{\rm high}$ & Cross section & Statistical uncertainty$^a$ \\
(keV) & (keV) & (barn) & (\%) \\
\hline
88.210 &  92.404  &  0.059  &  9 \\
92.404 &  96.748  &  0.059  &  5 \\
96.748 &  101.406 &  0.058  & 11 \\
101.406 &  106.408 &  0.057 &  8 \\
106.408 &  111.790 &  0.057 &  7 \\
111.790 &  117.591 &  0.056 &  8 \\
117.591 &  123.855 &  0.056 &  7 \\
123.855 &  130.634 &  0.055 &  7 \\
130.634 &  137.985 &  0.054 &  6 \\
137.985 &  145.974 &  0.054 &  6 \\
145.974 &  154.678 &  0.053 &  6 \\
154.678 &  164.185 &  0.053 &  7 \\
164.185 &  174.596 &  0.052 &  7 \\
174.596 &  186.030 &  0.051 &  6 \\
186.030 &  198.625 &  0.051 &  5 \\
198.625 &  212.544 &  0.050 &  5 \\
212.544 &  227.981 &  0.049 &  5 \\
227.981 &  245.162 &  0.049 &  5 \\
245.162 &  264.363 &  0.048 &  4 \\
264.363 &  285.911 &  0.047 &  4 \\
285.911 &  310.207 &  0.046 &  4 \\
310.207 &  337.739 &  0.046 &  4 \\
337.739 &  369.107 &  0.045 &  4 \\
369.107 &  405.060 &  0.044 &  4 \\
405.060 &  443.512 &  0.043 &  3 \\
\end{tabular}
\end{ruledtabular}
$^a$This value has to be added in quadrature with 
the overall systematic uncertainty of 10\%. 
\end{table}

\section{\label{sec:implications}Implications for the 
$s$-process abundance of the Pb/Bi isotopes}
  
Since $^{204}$Pb is shielded from the $r$ process by $^{204}$Hg, 
the observed solar abundance of $^{204}$Pb is only produced by 
the $s$-process branching at $^{204}$Tl, which is very sensitive 
to stellar temperature. Furtheron, the abundance of $^{204}$Pb is not
affected by the $\alpha$-recycling at the end of the $s$-process path (see
Sec.~\ref{sec:intro}), nor by the radiogenic contribution due to the decay of the
long lived U/Th isotopes.
Hence, the $^{204}$Pb abundance is determined by the strong temperature and
neutron density variations characteristic of the thermal pulses in
\textsc{agb} stars.

\begin{figure}[h]
\includegraphics[width=0.45\textwidth]{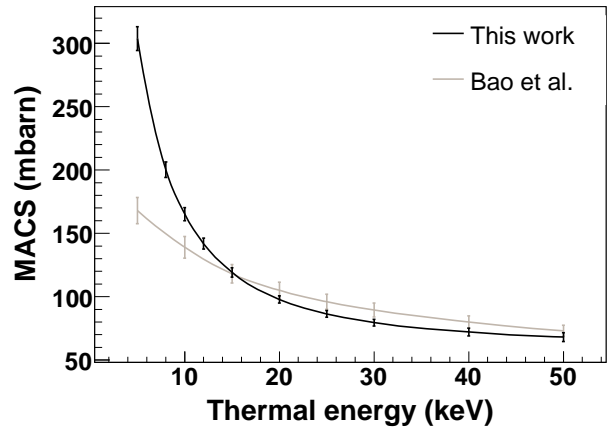}
\caption{\label{fig:macs} Maxwellian averaged cross section for 
$^{204}$Pb compared with data from Ref.~\cite{bao00}.}
\end{figure}

The capture cross section measured in this work was convoluted 
with a Maxwell-Boltzmann distribution in order to determine the 
Maxwellian averaged cross section (MACS) versus thermal energy, 
which is the relevant input quantity for nucleosynthesis calculations. 
The MACSs obtained in the present work are compared in 
Fig.~\ref{fig:macs} with the values reported in Ref.~\cite{bao00}, 
which are based on the only previous capture measurement
\cite{hor84,all73}. The large discrepancy of almost a factor of 
two below $kT=15$~keV is due to the resonances below $E_n = 2.5$~keV, 
which had not been reported before. At higher thermal energies 
the two data sets are in better agreement. Nevertheless, the 
present results are consistently smaller and about a factor of 
two more accurate. About 20\% of the MACS at 30~keV is due to 
the contribution of the average capture cross section beyond 
100~keV, reported in Table~\ref{tab:URR}.

\begin{table}[htbp]
\caption{\label{tab:macs} Maxwellian averaged cross section for
  $^{204}$Pb.}
\begin{ruledtabular}
\begin{tabular}{cc}
Thermal energy $kT$ & MACS \\
(keV) & (mbarn)\\
\hline
  5   &    304(9)\\
  8   &    200(6)\\
  10  &    165(5)\\
  12  &    142(4)\\
  15  &    119(4)\\
  20  &    98(3)\\
  25  &    86(3)\\
  30  &    79(3)\\
  40  &    72(3)\\
  50  &    68(3)\\
\end{tabular}
\end{ruledtabular}
\end{table}

The impact of the new MACS in the determination of the $s$-process 
abundances $N_s$ was estimated using the stellar model described
in Ref.~\cite{arl99}. Calculation have been made for stellar masses
of $M=1.5M_{\odot}$ and $3M_{\odot}$, and for a combination of
metallicities, [Fe/H]=$-0.3$ and [Fe/H]=$-1.3$, which have been shown
to account for the main and strong $s$-process components, 
respectively~\cite{tra99,tra01}. In spite of the much larger 
MACS at lower stellar temperature, the calculation based on the new 
cross section yields only a 4.6\% lower $s$-process production of
$^{204}$Pb, when compared to the same calculation made with the 
MACS of Ref.~\cite{bao00}. This result clearly illustrates that 
the production of $^{204}$Pb is mostly efficient at the higher
temperatures during He-shell flashes, when the decay of $^{204}$Tl 
is strongly enhanced \cite{TaY87}. 

The present estimate for the $s$-process abundance of $^{204}$Pb 
at the epoch of solar system formation is 95\% (relative to 
$^{150}$Sm). The uncertainty on the solar abundance of lead is
as high as 7.8\% according to Anders and Grevesse~\cite{and89}, rounded to
10\% by Lodders~\cite{lod03}. Within this uncertainty, which applies entirely to the solar
$s$-process contribution of $^{204}$Pb, the $s$-process abundance of
$^{204}$Pb obtained here is in perfect agreement with the expected value of 100\%.

A more consistent result will be attempted in a
comprehensive study~\cite{bis06} based on more stellar detailed model calculations
and on a complete set of new cross sections in the Pb/Bi region, e.g.
recent data for $^{207}$Pb~\cite{dom06b} and $^{209}$Bi~\cite{dom06} and new data for $^{206}$Pb.

\section{\label{sec:summary} Summary}

The neutron capture cross section of $^{204}$Pb has been measured in 
a high resolution time-of-flight experiment at the CERN n\_TOF 
facility. Data were obtained in the neutron energy range from 1~eV to 
440~keV. From a resonance analysis with the R-matrix code \textsc{SAMMY}
the capture widths of 170 resonances could be determined between 400~eV 
and 100~keV with an overall systematic uncertainty of 3\%. The average 
capture cross section in the energy interval from 100 to 440~keV was 
determined with an uncertainty of $\sim$10\%. From these results, 
Maxwellian averaged cross sections have been derived, which exhibit large 
discrepancies with respect to previous data. At thermal energies below 
$kT=$15~keV the present values are larger by up to a factor of two because 
new low-energy resonances could be included, whereas they are systematically 
lower by about 10\% at high values of $kT$, presumably because the 
neutron sensitivity of the older data had been underestimated. In any 
case, the systematic uncertainties could be improved by a factor of 
two as well. In spite of the significantly higher stellar cross sections 
at low $kT$, stellar model calculations show that the $^{204}$Pb 
abundance is not affected by more than 5\%. This result indicates that 
the production of $^{204}$Pb takes place during He-shell flashes, where 
the cross section differences with respect to the previous measurement 
are smaller and where the comparably high temperatures lead to an enhancement 
in the $\beta$-decay rate of $^{204}$Tl, thus favoring the $s$-process 
path towards $^{204}$Pb.

\begin{acknowledgments}
This work was supported by the European Commission (FIKW-CT-2000-00107), 
by the Spanish Ministry of Science and Technology (FPA2001-0144-C05), 
and partly by the Italian MIUR-FIRB grant "The astrophysical origin 
of the heavy elements beyond Fe". 
\end{acknowledgments}

\newpage 
\bibliography{article204Pb}

\end{document}